\documentclass[%
 reprint,
 amsmath,amssymb,
 aps,
]{revtex4-1}

\usepackage{graphicx}
\usepackage{dcolumn}
\usepackage{bm}

\begin{document}

\preprint{APS/123-QED}


\title{\boldmath
Connecting Electroweak Symmetry Breaking and Flavor: \\
A Light Dilaton ${\cal D}$ and a Sequential Heavy Quark Doublet $Q$}

\author{Wei-Shu Hou} 
 \affiliation{
Department of Physics, National Taiwan University,
 Taipei 10617, Taiwan
}


\begin{abstract}
The 125 GeV boson is quite consistent with the Higgs boson of the Standard Model (SM),
but there is a challenge from Anderson whether this particle is in the Lagrangian.
As LHC Run 2 
takes its final year of running, 
we ought to reflect and make sure
we have gotten everything right. 
The ATLAS and CMS combined Run 1 analysis claims 5.4$\sigma$
measurement of vector boson fusion (VBF) production
that is consistent with SM, which seemingly refutes Anderson.
However, to verify the source of electroweak symmetry breaking (EWSB),
we caution that VBF measurement is too important
for us to be imprudent in any way,
and gluon-gluon fusion (ggF) with similar tag jets must be simultaneously \emph{measured}, 
which should be achievable at LHC Run 2.
The point is to truly test the dilaton possibility,
the pseudo-Goldstone boson of scale invariance violation.
We illustrate EWSB by
dynamical mass generation of a sequential quark doublet $Q$
via its ultrastrong Yukawa coupling, and argue
how this might be consistent with a 125 GeV dilaton, ${\cal D}$.
The ultraheavy $2m_Q \gtrsim 4$--5 TeV scale explains
the absence of New Physics so far, while the mass generation
mechanism shields us from the UV theory for the strong Yukawa coupling.
Collider and flavor physics implications are briefly touched upon.
Current Run 2 analyses show correlations between the ggF and VBF measurements,
but the newly observed $t\bar tH$ production at LHC poses a challenge.

\begin{description}
\item[PACS numbers]
11.15.Ex	
12.15.Ff	
14.65.Jk 
14.80.-j	
\end{description}
\end{abstract}

\pacs{Valid PACS appear here}
\maketitle


\section{Higgs, Anderson, and 
all that
}
%
Spontaneous symmetry breaking (SSB) was introduced into particle physics
by Nambu as cross-fertilization from superconductivity (SC).
In an explicit model with Jona-Lasinio (NJL), Nambu illustrated~\cite{Nambu:1961tp}
how the nucleon mass $m_N$ could arise from dynamical chiral symmetry breaking (D$\chi$SB),
with the pion emerging as a pseudo-Nambu-Goldstone (NG) boson.
Subsequent work lead to the BEH mechanism~\cite{BE, H} of electroweak symmetry breaking (EWSB),
which became~\cite{Weinberg:1967tq, Salam:1968rm} part of the Standard Model (SM).
The recently discovered 125 GeV boson~\cite{Higgs2012} seems consistent with
the Higgs boson of SM by every count.
This has in turn stimulated condensed matter physicists
to pursue their own ``Higgs'' mode.

A ``Higgs'' mode was recently observed~\cite{squalid-H}
in disordered SC films near the SC-insulator
quantum critical point, far below the $2\Delta$ double-gap threshold.
Here, $\Delta$ is the ``energy gap'' of the SC phase,
which was maintained throughout the experiment.
This ``light Higgs'' mode contrasts with ``amplitude modes''
around $2\Delta$ that were claimed long ago~\cite{LittleVarma}.
%
%
Anderson, who originated the nonrelativistic version of
the BEH mechanism, 
praised~\cite{Anderson} Nambu for elucidating~\cite{Nambu:1961tp}
the dynamical generation of $m_N$, a ``{mass gap}'',
by drawing analogy with SC:
a scalar boson in NJL-type of models has mass $\sim 2m_N$
is an ``amplitude mode''.
%
%
Anderson then turned to challenge particle physics~\cite{Anderson}:
``\emph{If superconductivity does not require an explicit Higgs in the Hamiltonian
to observe a Higgs mode, might the same be true for the 126 GeV mode?}",
hence jesting ``\emph{Maybe the Higgs boson is fictitious!}".
He then stressed the importance of Ref.~\cite{squalid-H}, as
``it bears on \emph{the nature of the Lagrangian} of the Standard Model''.

As Anderson coined the word ``emergent''~\cite{Anderson:1972pca}
for phenomena that are not inherent in the Lagrangian,
he challenges the \emph{elementary} nature of the 125 GeV boson.

What do we really know about the 125 GeV boson? If it is not
the Higgs boson $H$ of SM, then what else could it be?
In this paper, we revamp the idea that the observed boson
could still be a dilaton ${\cal D}$ from spontaneous scale invariance violation.
We argue that this can be truly excluded only by data-based
simultaneous measurement of both the vector boson fusion (VBF) process
and gluon-gluon fusion (ggF) plus similar tag jets.
This is hopefully achievable with Run 2 data
at the Large Hadron Collider (LHC),
despite the existing claim~\cite{H-combo} already with Run~1 data.
We then elucidate how EWSB might arise from
\emph{dynamical} mass generation of a sequential quark doublet $Q$
through its ultrastrong Yukawa coupling,
resulting in $2m_Q$ that is far above 125 GeV,
which echoes the result of Ref.~\cite{squalid-H}.
One should, of course, avoid directly matching a dilaton
to the ``Higgs'' mode of Ref.~\cite{squalid-H}.

\begin{figure}[b!]
{
 \includegraphics[width=75mm]{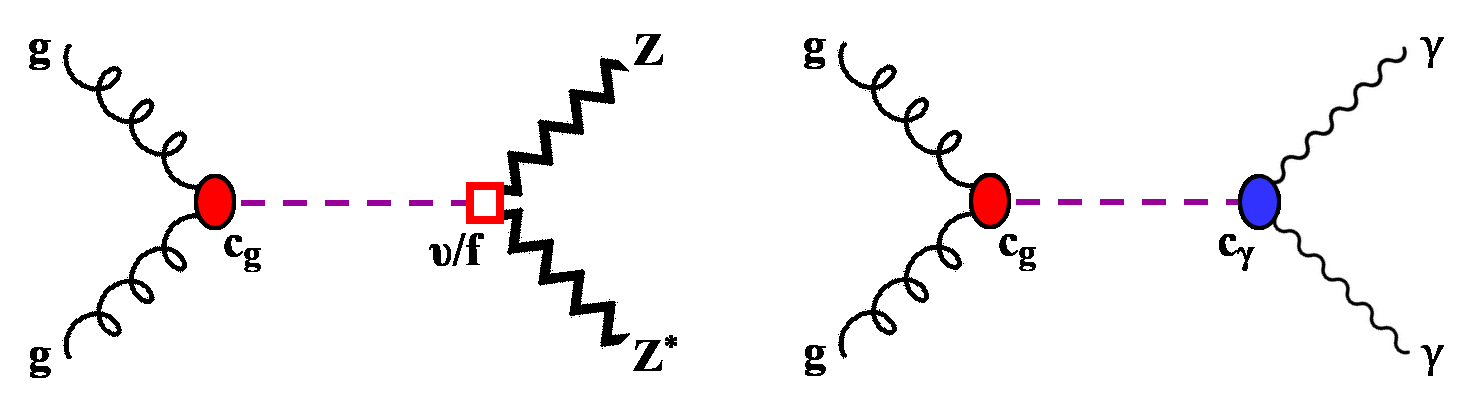}
}
\caption{
Gluon-gluon fusion production and $ZZ^*$, $\gamma\gamma$ decay
of the 125 GeV boson (dashed line).
} \label{fig:gg}
\end{figure}
%


%

The discovery of the 125 GeV boson is perceived as
due to ggF production and driven by
$ZZ^*$ (two lepton pairs) and $\gamma\gamma$ channels, as illustrated in Fig.~1.
It is usual to measure the relative strength with respect to (w.r.t.) SM,
$\mu \equiv (\sigma\cdot {\cal B})/(\sigma\cdot {\cal B})|_{\rm SM}$
(i.e. 1 for $H$ of SM).
In terms of the coefficients $c_g$, $\hat v \equiv v/f \equiv 1/\hat f$
(meaning of $f$ defined later) and $c_\gamma$, as illustrated in Fig.~1,
of the $gg$, $ZZ$ and $\gamma\gamma$ couplings of the 125 GeV boson
w.r.t. SM, what ATLAS+CMS observe~\cite{H-combo} is
\begin{align}
  \mu_{ZZ^*}
   & \cong \frac{c_g^2\, \hat v^2}
              {0.910\,\hat v^2 + 0.0868\,c_g^2} \simeq \; 1, \\ 
  \mu_{\gamma\gamma}
   & \cong \frac{c_g^2\, c_\gamma^2}
              {0.910\,\hat v^2 + 0.0868\,c_g^2}  \gtrsim \; 1, 
\end{align}
where we assume $v/f$ applies also to fermions, 
ignored the tiny $\gamma\gamma$ and $Z\gamma$ decays in the denominator,
and assume the absence of invisible width $\Gamma_{\rm inv}$.
We have taken nominal values of SM decay rates
for $VV^*$ and $f\bar f$ final states, and used $\Gamma_{\rm SM} \simeq 4$ MeV.
Thus, the denominator in above equations is nothing but $\Gamma/\Gamma_{\rm SM}$.
%
%

Eq.~(1) is of course satisfied by the SM case of $c_g$, $\hat v \simeq 1$.
But if one allows $c_g > 1$, then the allowed
value for $1/\hat v$ increases, which onsets quickly
(hence width $\Gamma$ drops first as $c_g$ increases from 1,
before picking up for large $c_g$),
but saturates to
$f/v \simeq 1/\sqrt{0.0868} \simeq 3.394$ as $c_g \to \infty$,
which can be seen easily from Eq.~(1).
For example, $f/v \simeq 2$, 3, 3.22, 3.33, respectively,
for $c_g \simeq 1.18$, 2.04, 3.0, 4.93.
The mild inequality of Eq.~(2) is easier to satisfy.
Besides $c_g,\;\hat v \simeq 1$, for the aforementioned values of
$(c_g,\;\hat f) \simeq (1.18,\; 2)$, (2.04, 3), (3.0, 3.22), (4.93, 3.33),
one has $|c_\gamma| \gtrsim 0.50$, 0.333, 0.311, 0.30, respectively,
reaching the asymptotic $\sqrt{0.0868} \simeq0.295$ for very large $c_g$.
These examples for $c_g$, $v/f$ and $|c_\gamma|$ came
as a result of Higgs width and branching ratio considerations.

For large $c_g$ and with $VV^*$ and $f\bar f$ rates suppressed by $v/f$,
the predominant decay would be the $gg$ mode, just as in production.
%
%
New Physics could affect allowed $c_g$, $v/f$ and $c_\gamma$ values,
but just the presence of $\Gamma_{\rm inv}$ would only make matters worse,
as it would disallow a compensating effect of smaller $\hat v$.

Measurements are remarkably consistent with SM, but
one should probe individual coefficients directly.
If the 125 GeV boson is a dilaton ${\cal D}$,
the (pseudo-)NG boson from SSB of scale invariance,
then $c_g$ and $c_\gamma$ are determined by the trace anomaly
of the energy momentum tensor, which would depend on the beta functions
of QCD and QED, respectively,
while $v/f$ is a common factor mentioned
by Altarelli~\cite{Altarelli:2013lla} as late as 2013:
%
``The Higgs couplings are proportional to masses: a striking signature ...'',
but ``also true for a dilaton, up to a common factor''.
Thus, $f$ is the dilaton decay constant.

That a dilaton could be confused for a light SM Higgs boson
was stressed by Ref.~\cite{Goldberger:2008zz} in 2008,
before the advent of LHC. However, the example given was to
have QCD and QED ``embedded in the conformal sector at high scale'',
hence $c_\gamma = -17/9$, and $c_g = 11 -2N_{\rm light}/3  = 23/3$,
a case (and similar large values) that is definitely ruled out~\cite{no-dil},
causing many to write-off the dilaton.
In fact, early papers~\cite{Chacko,Csaki} on dilaton interpretation of
the new 125 GeV boson noted that data preferred ``Higgs-like''
dilaton of $f \simeq v$, which is not what we advocate.
For example, starting from the $c_g$, $c_\gamma$ and $v/f$ parametrization,
Ref.~\cite{Csaki} showed that $v/f \sim 1/3$ was ruled out already by
early Run 1 data. On closer inspection, however, the authors of
Ref.~\cite{Csaki} have $c_g$, $c_\gamma$ themselves scaled by $v/f$, which
is opposite the trend of large values of Ref.~\cite{Goldberger:2008zz},
and we are not certain of the full generality.
In view of the Anderson challenge, the dilaton should be
kept in mind and tested without prejudice, to the purist
criteria of Elander and Piai~\cite{Elander:2012fk}
of keeping $c_g$, $v/f$ and $c_\gamma$ as parameters.

\begin{figure}[t!]
\vspace{3mm}
{
 \includegraphics[width=40mm]{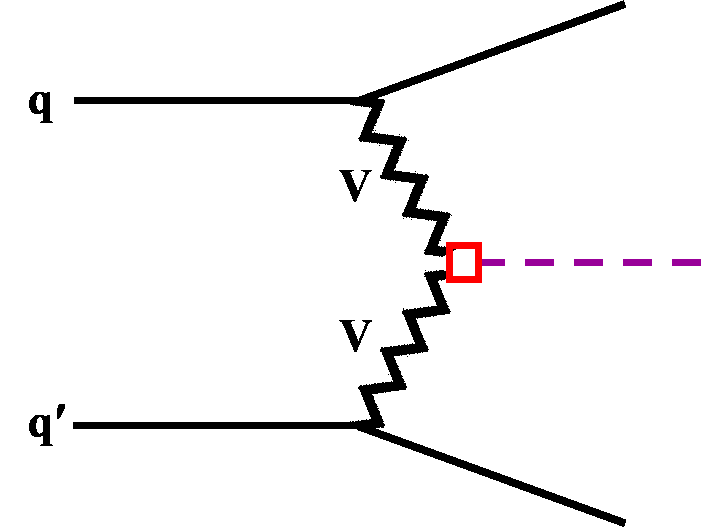}
}
\caption{
Vector boson fusion production of 125 GeV boson. 
} \label{fig:VBF}
\end{figure}
%


One might say that the $VV$ coupling has already been
measured with Run 1 data: the combined analyses of
the ATLAS and CMS experiments together claim~\cite{H-combo}
$5.4\sigma$ measurement of VBF production,
finding consistency with SM hence $v/f \sim 1$,
which would run against the dilaton possibility.
It is certainly true, and very important, that the $VV$ coupling of the
125 GeV boson can be probed directly by the VBF process, as illustrated in Fig.~2.
In the following section, we begin with a critique of this 5.4$\sigma$ measurement,
cautioning that it may still be premature.
We update with Run 2 results that have become available,
but defer it to Sec.~V.

In this paper, we take the experimentally observed 125 GeV boson as
the dilaton ${\cal D}$, without accounting for its true origins.
We revamp the case for dynamical EWSB
by ultrastrong Yukawa coupling of a sequential quark doublet $Q$,
and elucidate why it might be consistent with the emergence of a dilaton.
The approach shields one from the high energy completion
behind this strong Yukawa coupling,
including the origin of scale invariance breaking,
hence the emergence of the dilaton itself.

\section{
On Observing VBF 
}

We first note that the Run 1 VBF measurements by ATLAS and CMS
are not individually significant yet,
as the cross section is $\sim 1/12$ of the leading ggF process in SM.
Combining datasets, when analyses are still limited by statistics,
is suitably common.
However, the combined analysis of LHC Run 1 data by ATLAS and CMS,
claiming 5.4$\sigma$ measurement of VBF process, has some weaknesses.
We offer here some simple critique,
with discussion of Run 2 situation deferred to Sec.~V.

First, an issue of semantics.
Recall the usage of Higgs-\emph{like}
for the 125 GeV boson up to early 2013.
By same token, in Run 1 one is really probing
VBF-\emph{like} production, rather than genuine VBF.
This is because it is based on multivariate analysis of categorized data~\cite{H-combo}.
As radiation of vector boson $V$ is rather analogous to synchrotron radiation,
it is effective only when each ``spent'' quark retains most of
the initial parton momentum.
But since $m_V$ is sizable, genuine VBF requires
\emph{two ultra-energetic tag jets} that are 
necessarily back-to-back~\cite{farforward}
with large $m_{j_1j_2}$ and large rapidity separation,
and little color radiation in the rapidity gap.
The categorized analysis is a compromise due to limited statistics.
If statistics were sufficient, one would always cross-check
with a high purity VBF selection (cut-based analysis)
that would beat ggF background down to a true minimum.

Second, with ggF production the leading process, one needs to 
explore analysis methods to simultaneously measure both VBF 
and ggF production with similar tag jets, 
methods that require statistical power to achieve the separation.
The current VBF measurement relies on predicting the 
jet-tagged ggF yield in the two-jet (VBF-like) category, ggF+jj,
and subtracting it from the measured yield~\cite{mired}.
Although this extrapolation relies on Monte Carlo, 
experimentally the MC predictions for the 0-, 1- and 2-jet 
categories are checked with data, and the systematic uncertainty 
of extrapolation, though not small, is under control. 
As integrated luminosity accumulates, the separation power 
between VBF and ggF+jj will improve and eventually lead to
a systematic error on VBF that is lower than 
the one provided by the current subtraction method. 
In this case, biases coming from the ggF side will be removed, 
and systematics will be due to the level of control of 
ggF and VBF kinematic distributions.

%

%
Third, the prominence of ``Higgs boson''
discovery means bias necessarily seeps
into the analyses, especially after the 2013 Nobel prize.
But there is no good way to combine potential biases~\cite{bias}.
Finally, the $5\sigma$ claim has the connotation
that \emph{observation} is achieved.
But identifying the true source of EWSB is too important an issue
to not keep the highest standards.

We advocate that one should await verdict
on VBF from the much larger dataset
that is already two-thirds its way through at LHC Run 2.
Note that, despite some hints for $t\bar tH$ production
in both Run 1~\cite{H-combo} and early 13 TeV data~\cite{ICHEP2016},
they are less significant.
We turn to a brief survey of currently available Run 2 results in Sec.~V.

In view of Anderson's challenge,
we take the 125 GeV boson as an emergent dilaton,
and turn to recount how a new sequential quark doublet $Q$
could self-generate $m_Q$ by its ultrastrong Yukawa coupling.
This dynamical EWSB mechanism \emph{may allow} a dilaton to emerge,
but does not quite explain it.
%

The four generation (4G) model was supposedly
``killed by the Higgs discovery''~\cite{Stone:2012yr},
because adding $t^\prime$, $b^\prime$ to $t$
in the triangle loop for $ggH$ coupling would
enhance the amplitude by $\sim 3$, 
hence the cross section by 9, which is not observed~\cite{Higgs2012}.
But, there is {nothing really wrong} with 4G quarks,
\emph{except} {this ``Higgs'' cross section},
which could be for a dilaton, as we have just stressed.
As already commented, $c_g \gtrsim 3$, compensated by $v/f \sim 1/3$,
with appropriate $c_\gamma$, also gives
$\mu_{ZZ^*} \sim 1$ and $\mu_{\gamma\gamma}\gtrsim 1$. 
%
%


\section{The Yukawa Coupling Enigma}
%

Yukawa couplings of fermions are an enigma,
but an elementary Higgs field is not needed to define them.
There is a {dynamical} difference between electroweak (EW) theory
vs. QED and QCD, where decoupling~\cite{Appelquist:1974tg} is the rule.
Nondecoupling of heavy quarks in EW processes,
such as EW penguin effects in $b \to s\ell^+\ell^-$~\cite{Hou:1986ug},
is rooted in the {Yukawa coupling}, which grows with mass.

As this author learned particle physics, SM began to enter textbooks,
so the Lagrangian was taken for granted.
%
The SM Lagrangian has a built-in complex scalar doublet,
and it was Weinberg who introduced~\cite{Weinberg:1967tq}
the Yukawa coupling for fermion mass generation.

By time of LHC turn-on, however, the weak vertex
\begin{equation}
\frac{1}{\sqrt{2}}\, gV_{ij}\, \bar u_i \gamma_\mu L d_j\, W^\mu,
\end{equation}
had become 
firmly established by LEP and B factory data.
Since all particles in Eq.~(3) are massive, and since 
the longitudinal $W_L$
propagates by the $\frac{k_\mu k_\nu}{M_W^2}$ factor,
replacing $W^\mu$ by $\frac{k_\mu}{M_W}$ in Eq.~(3)
and using the Dirac equation, one gets~\cite{Hou:2012az}
\begin{equation}
\frac{1}{\sqrt{2}}\, V_{ij}\,
\bar u_i \left(\lambda_i L - \lambda_j R\right) d_j\, G.
\end{equation}
The weak coupling $g$ cancels against $M_W = \frac{1}{2}gv$, and
\begin{equation}
\frac{\lambda_Q}{\sqrt{2}} \equiv \frac{m_Q}{v},
\end{equation}
is exactly the Yukawa coupling of the NG boson $G$,
with both left- and right- chiral couplings
emerging from a purely left-handed vector coupling!
{The point}
{is: \emph{no Lagrangian is used}},
hence Yukawa couplings are 
experimentally established,
and the longitudinal $W_L$ is the ``eaten'' NG boson,
{without touching upon whether there is an {elementary}
Higgs boson} {or field}.

One may say that the above is nothing but the Goldstone theorem~\cite{Goldstone}.
What we have elucidated is that all our knowledge of Yukawa couplings,
including CKM matrix elements $V_{ij}$ and the unitarity of $V$,
are extracted through their dynamical, nondecoupling, effects.
They arise from the NG bosons, without reference to an
elementary Higgs doublet field, nor its remnant particle.

Anderson's point, then, is that we need to
make sure the 125 GeV boson is in fact the remnant of
a complex scalar doublet in the SM Lagrangian,
as we have discussed in previous section.

But Yukawa couplings are truly an enigma:
\emph{we know not} what determines their values that range from
$\lambda_{u,\, d}\sim 10^{-5}$ to $\lambda_t \cong 1$,
while modulated by $V_{ij}$ that exhibit hierarchical pattern,
they are the sources of all known flavor physics and $CP$ violation (CPV).
With quark Yukawa couplings spanning 5 orders already,
we now argue that raising by another order to the ``extremum''
value of $\lambda_Q \gtrsim 4\pi$, it could induce dynamical EWSB.

\begin{figure}[t!]
{
 \includegraphics[width=80mm]{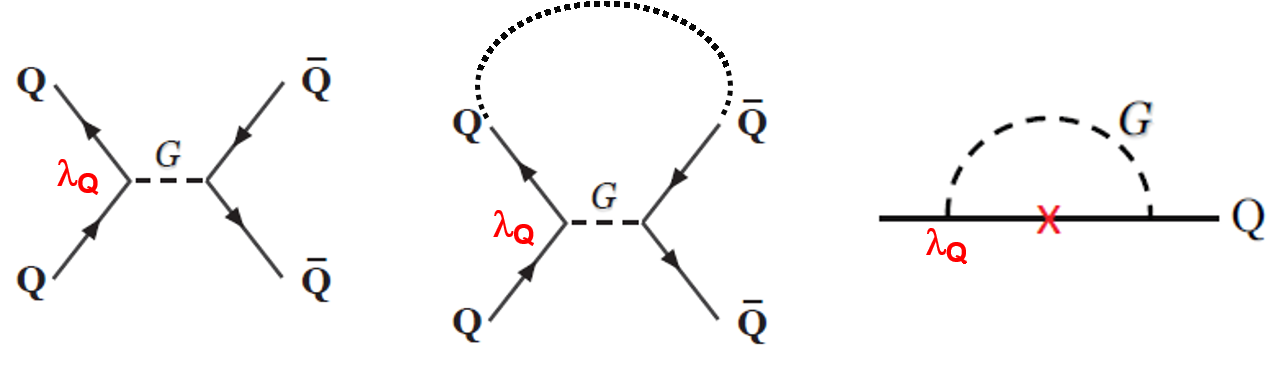}
}
\caption{
(left) $Q\bar Q \to Q\bar Q$ scattering by exchange of NG boson $G$ (or longitudinal $V_L$);
(center) connecting $Q$ to $\bar Q$ across exchanged $G$;
(right) self-energy of $Q$ by $G$ loop, with mass generation illustrated by cross ($\times$).
} \label{fig:ScatGap}
\end{figure}
%



\section{Ultrastrong Yukawa-induced EWSB and the Dilaton}

After restart of LHC, by 2010 search limits on $m_{b^\prime}$, $m_{t^\prime}$
rose quickly beyond the nominal ``unitarity bound''~\cite{Chanowitz} of $\sim 550$ GeV,
but search continued for unitarity bound violating (UBV) 4G quarks.
The heavy mass just implies very strong Yukawa coupling,
and EW precision observables $S$ and $T$ demand
\emph{nearly} degenerate~\cite{Kribs:2007nz, Holdom:2009rf} $t^\prime$-$b^\prime$,
hence we denote as $Q$.
Note that a small $m_{t^\prime}$--$m_{b^\prime}$ splitting is needed
to compensate~\cite{Kribs:2007nz} between 
$S$ and $T$ as $m_Q$ and Higgs mass both become very heavy.

UBV implies bad high energy (H.E.) behavior for
$Q\bar Q \to Q\bar Q$ scattering, which is
dominated by $G$ (i.e. 
 $V_L$) exchange, as shown in Fig.~3(left).
The range of interaction, $1/M_W$, becomes large compared with $1/m_Q$ for heavier $Q$.
This runs against the intuition for short distance or UV remedy
of the bad H.E. behavior, whether based on UBV or NJL folklore.
Linking~\cite{Hou:2012az} a $Q$ to a $\bar Q$ across the exchanged $G$,
Fig.~3(center), the $Q\bar Q \to Q\bar Q$ scattering turns into
the self-energy of $Q$, where the exchange momentum $q$ is summed over.
This becomes a ``gap equation'' for generation of $m_Q$, the ``mass gap'',
as illustrated in Fig.~3(right), with the cross ($\times$)
representing the self-energy function itself.
A nontrivial solution would mean mass generation.
As the chiral symmetry is the SU(2)$_{\rm L}$ gauge symmetry,
D${\chi}$SB means dynamical EWSB,
which is in reverse of Weinberg~\cite{Weinberg:1967tq}.

The self-energy in Fig.~3(right) differs from NJL~\cite{Nambu:1961tp},
which uses a dimension-6 four-quark operator that leads to
a closed ``bubble'' with freely running loop momentum $q$
but is independent of external momentum $p$,
{with cutoff $\Lambda$ provided by the operator coefficient}.
In contrast, 
the NG boson loop of Fig.~3(right) manifests the long-distance nature,
while the $QQG$ coupling brings the external momentum $p$ into the loop.
Thus, {the Yukawa-induced gap equation is different from NJL}
and more intricate.
{Note there is \emph{no scale parameter},
{as tree level $m_Q^0 = 0$ by gauge invariance}.

To formulate the gap equation for mathematical solution,
one needs to fix {the range of integration for $q$}.
%
With no new physics found up to 1 to several TeV by summer 2011,
the self-consistent and simplest ansatz~\cite{Hou:2012az}
is to integrate $q^2$ up to $(2m_Q)^2$,
such that the NG boson $G$ in the loop is justified.
%
By keeping $\lambda_Q$ defined in Eq.~(5) as a \emph{parameter},
the scale $v$ is brought in to make contact with experiment.

With this ansatz of integration limit being twice the generated mass $m_Q$,
the gap equation was solved numerically~\cite{Mimura:2012vw} in the ladder approximation.
Despite the urge to keep $m_Q$ below TeV for sake of LHC phenomenology, 
a nontrivial solution demanded
\begin{equation}
\lambda_Q \gtrsim 4\pi,\ \ \ \ (m_Q \gtrsim {\rm 2\ TeV!})
\end{equation}
i.e. at ``Naive Dimensional Analysis'' (NDA)
strong coupling~\cite{Manohar:1983md} of $4\pi$ or higher~\cite{int_range}.
D$\chi$SB, hence dynamical EWSB, can occur at ``extremum'' coupling strength!
Shortly after submission, however, the 125 GeV boson was announced~\cite{Higgs2012},
so it took one and half years to get the work published~\cite{Mimura:2012vw},
which was largely ignored.

The challenge from Anderson~\cite{Anderson}, however,
throws a different light.
In the Yukawa-dynamical EWSB, the self-energy
sums over $Q\bar Q \to Q\bar Q$ scattering,
hence is a \emph{pairing mechanism},
much like Cooper pairs of BCS theory of superconductivity,
which NJL tried to emulate~\cite{Nambu:1961tp}.
We have already expounded the difference with NJL,
and the numerical solution suggests EWSB occurs at
NDA-strong $4\pi$ strength, hence
perturbation has broken down absolutely~\cite{Manohar:1983md}.
%
For $\lambda_Q$ consistent with Eq.~(6),
$m_Q$ is generated, which means a $Q\bar Q$ condensate has formed,
hence the exactly massless 
NG boson $G$ is in fact a \emph{\emph{$Q\bar Q$ boundstate}}.
All these can be viewed from the perspective of
$Q\bar Q$ scattering in the massive world~\cite{UR-boundstate}.
This dynamical {mechanism can induce EWSB, without ever having an}
{``explicit'' Higgs in the Lagrangian}. 
{And much like NJL} model, there should be ``amplitude'' modes,
such as scalar bosons, around $2 m_Q \sim 4$--$5$ TeV.

We did not~\cite{Mimura:2012vw}, however, anticipate
a light boson far below $2 m_Q$,
but {a light 125 GeV boson} \emph{emerged}.
In face of the challenge by Anderson, we take it to be a dilaton~\cite{dilaton}.
But how does it make sense in context?

{Recall that our gap equation}
{based on Yukawa coupling $\lambda_Q$ has no scale},
and contact with $v$ was introduced self-consistently
by ansatz of integration up to $2m_Q = \sqrt2\lambda_Q v$. 
Nontrivial numerical solution to our no-scale formulation,
hence $m_Q$ generation, would also seemingly break scale invariance.
This may \emph{allow} a dilaton $\cal D$ to emerge~\cite{dilaton2},
but we neither predicted it,
nor do we know how $m_{\cal D}$ is generated.
The dilaton should arise from the true origin of
scale invariance violation, which we
conjecture to be the theory of strong Yukawa coupling that explains Eqs.~(5) and (6).
As elucidated in Sec.~III, Yukawa couplings arise empirically from EW physics,
without the need for a Higgs field to define them.
{Note that $Q\bar Q$ condensation and the integration limit of}
{$2m_Q$, shield us from the actual UV theory},
which is likely not far beyond the rather high $2m_Q$
(because of the strong $\lambda_Q$).
We do not know what it is, except that it is strongly coupled,
and likely conformal~\cite{Goldberger:2008zz, Chacko, Csaki}.

So, we have New Physics both within and beyond SM.
Rather than the Higgs field, the agent of mass, or EWSB, is
$Q\bar Q$ condensation via its own ultrastrong $\lambda_Q$.
The 125 GeV boson is a dilaton ${\cal D}$ that descends from some unknown UV sector;
unlike the NG boson $G$, it cannot be a pure $Q\bar Q$ boundstate.

\section{ LHC Run 2 Results}

The original version of this essay was prepared around the time
of the ATLAS and CMS Run 1 combined analaysis~\cite{H-combo}.
Since then, some Run 2 results (13 TeV) have become available,
and it is necessary to check how our discussion so far
survives data scrutiny. We will see that our scenario remains
potent, and in fact ggF vs VBF production measurements do
show some ``symptoms''. However, the observation,
by both ATLAS and CMS, of $t\bar tH$ production poses a challenge.

Without quoting detailed errors, and often dropping
insignificant results without comment,
let us give a brief survey of what is currently available:
\begin{itemize}
\item {\boldmath $ZZ^*$}:
 Both experiments have made available the analyses up to 2017 data,
 at 79.8 fb$^{-1}$ and 77.4 fb$^{-1}$, respectively, for ATLAS and CMS.

 For ATLAS~\cite{ZZ-2-ATL}, while $\mu_{\rm ggF} \simeq 1$ is measured,
      the $\mu_{\rm VBF} \simeq 2.8$ value is rather large.

 For CMS~\cite{ZZ-2-CMS}, $\mu_{\rm ggF,b\bar bH} = 1.15^{+0.18}_{-0.16}$ is measured,
      $\mu_{\rm VBF} = 0.69^{+0.75}_{-0.57}$ is barely 1$\sigma$,
      reflecting in part the absence of events in 2016 data (36.9 fb$^{-1}$).

 Could these ``fluctuations'' reflect a much larger ggF production rate,
  but with an analysis strategy centered around SM expectation?
%
\item {\boldmath $\gamma\gamma$}:
 Results are for 36.1 fb$^{-1}$ and 35.9 fb$^{-1}$, respectively
 (i.e. 2016 data), for ATLAS and CMS.

 For ATLAS~\cite{gaga-2-ATL},
      $\mu_{\rm ggF} = 0.81^{+0.19}_{-0.18}$ is mildly less than 1,
      but $\mu_{\rm VBF} = 2.0^{+0.6}_{-0.5}$ is again rather large.

 For CMS~\cite{gaga-2-CMS}, $\mu_{\rm ggF} = 1.10^{+0.20}_{-0.18}$ looks reasonable,
      but $\mu_{\rm VBF} = 0.8^{+0.6}_{-0.5}$ is not inconsistent with zero.

 The trend between ATLAS and CMS are again opposite.
 In addition to the possibility that ggF production could be much stronger
  than assumed, it may reflect difference in analysis choice(s).
\item {\boldmath $WW^*$}:
 Both experiments are only for 2016 data.

 For ATLAS~\cite{WW-2-ATL}, the measured $\sigma\cdot{\cal B}|_{\rm ggF}$ at 6.3$\sigma$
       is $\simeq 20\%$ larger than SM expectation,
      while $\sigma\cdot{\cal B}|_{\rm VBF}$ is found at 1.9$\sigma$,
       w.r.t. SM expectation at 2.7$\sigma$.

 For CMS~\cite{WW-2-CMS}, $\mu_{\rm ggF} = 1.38^{+0.21}_{-0.24}$
       is 1.6$\sigma$ above SM,
      while ${\rm VBF} = 0.29^{+0.66}_{-0.29}$ is consistent with zero,
      reflecting in part the null result in 2016 data.
\item {\boldmath $\tau\tau$}:
 Based on 2016 data, ATLAS has recently joined CMS in claiming
  observation.
 Given the large backgrounds for $gg \to H \to \tau^+\tau^-$,
  the observation was made with ``jet assistance''. 

 For CMS~\cite{tata-2-CMS}, $\mu = 1.09^{+0.27}_{-0.26}$
  is at 4.9$\sigma$ (combining with Run 1 to become 5.9$\sigma$)
  which combines the 0-jet, Boosted, and VBF measurements.
  Not surprisingly, 0-jet is barely 1$\sigma$, so the measurement
  comes from the latter two. But our question of jet-tagged ggF vs VBF remains.

 For ATLAS~\cite{tata-2-ATL}, combining Boosted and VBF categories
  gives 4.4$\sigma$ (4.1$\sigma$), improving to 6.4$\sigma$ (5.4$\sigma$)
  when combined further with Run 1.
  The expected SM significance is given in parenthesis.
\item {\boldmath $b\bar b$}:
 Both ATLAS and CMS find evidence. 
 The large $b\bar b$ cross section from QCD implies
  jet-tag-assistance would not work, and measurements are
  based on $VH$ associated production, where both experiments
  use $VZ$ production for validation.
 Combining 2016 data with Run 1,
  ATLAS~\cite{bb-2-ATL} and CMS~\cite{bb-2-CMS} experiments
   find evidence at 3.6$\sigma$ (4.0$\sigma$)
   and 3.8$\sigma$ (3.8$\sigma$), respectively.
 Both experiments find excess events in $m_{b\bar b}$ above the $Z$ pole.
\item Combinations:
 CMS has put out a combination of analyses based on 2016 data,
  while ATLAS has combination of only $ZZ^*$ and $\gamma\gamma$ modes.

 For CMS~\cite{Combo-2-CMS}, $\mu_{\rm ggF} \simeq 1.23$ is about 1$\sigma$ above SM,
      while $\mu_{\rm VBF} \simeq 0.73$ is about 1$\sigma$ below.
  $WH$ is found large, about twice SM expectation, while
  $ZH$ is consistent with SM and 2$\sigma$ away from zero.

 For ATLAS~\cite{combo-2-ATL}, $\mu_{\rm ggF}$ is consistent with 1,
      but $\mu_{\rm VBF}$ is greater than 2 and is rather large.
  $VH$ is found consistent with zero.
\item {\boldmath $t\bar tH$}:
 By adding 2017 data for the $H\to ZZ^*$ and $\gamma\gamma$ modes,
  ATLAS has recently joined CMS in
  observation of $t\bar tH$ production at the LHC.

 For CMS~\cite{tth-CMS}, based on $H$ decay to the five modes of
  $WW^*$, $ZZ^*$, $\gamma\gamma$, $\tau^+\tau^-$ and $b\bar b$,
  and combining 2016 data with Run 1, the measurement of
  $\mu_{t\bar tH} = 1.26^{+0.31}_{-0.26}$ makes a 5.2$\sigma$ observation.

 ATLAS~\cite{tth-ATL} was a bit unlucky with similar data set.
  Adding 2017 data to $H\to ZZ^*$ and $\gamma\gamma$
  and combining with 2016 data for the other three modes
  give $\mu_{t\bar tH} = 1.32^{+0.28}_{-0.26}$,
  achieving 5.8$\sigma$ (4.9$\sigma$) observation with 13 TeV data alone.
  Combining with Run 1, the significance becomes 6.3$\sigma$ (5.1$\sigma$).
%
\end{itemize}

So, what to make of the available 13 TeV results?
For the two Run 1 drivers, $ZZ^*$ and $\gamma\gamma$,
there are some apparent correlation, or fluctuations,
between ATLAS and CMS results on ggF vs VBF.
While CMS shows a mild deficit for VBF,
ATLAS shows a rather large excess,
so perhaps our criticism given in Sec.~II may have some bearing.
As already raised above,
could this be due to a much larger ggF production rate,
compounded by analysis strategies based on SM mindset
(e.g. the underlying ggF Monte Carlo)?
Rather than genuine VBF with very energetic forward and backward jets,
perhaps what is measured is ggF with VBF-like double tag-jets.

The observation of $H \to \tau^+\tau^-$ is of significance for
direct evidence of $H$ coupling to fermions.
However, given the nature of jet-assistance, our criticism remains,
that one cannot be sure that it is not actually
jet-tag-assisted ggF production that is measured.
For $WW^*$, both experiments seem to have observed ggF production,
but the indication for VBF is quite weak.
Without a mass bump, the analysis is quite different
from $ZZ^*$ and $\gamma\gamma$, so our point of
simultaneous measurement of ggF with VBF-like jet-tags
and genuine VBF remains valid.

What may be a little worrisome is the evidence,
from both ATLAS and CMS, for $b\bar b$ in $VH$ production.
The production is quite distinct from ggF,
and should be suppressed if $v/f$ is of order 1/3 or so,
but there is clear excess in $m_{b\bar b}$ above $M_Z$,
which we cannot explain.

The observation of $t\bar tH$ production,
with mild excess above SM for both ATLAS and CMS,
could be devastating to our proposed scenario.
The process is nothing but $H$ radiating off the
QCD production of a $t\bar t$ pair,
hence a direct measurement of $ttH$ coupling,
and consistency with SM expectation offers genuine support
for the top Yukawa coupling, $\lambda_t \cong 1$.
The process ought to be suppressed by the universal factor
$v^2/f^2$ for the dilaton case.
Even if a greatly enhanced $c_g$ leads to
125 GeV boson emission off a gluon line
for QCD production of $t\bar t$ pair,
that it would also mimic $t\bar tH$ production strength
in SM would seem rather contrived.

In lieu of plainly accepting defeat,
we do caution that Yukawa couplings are a true enigma,
and whether there are 4G quarks or not,
the top quark is special.
As a reminder, Yukawa couplings
are hidden by spontaneous (dynamical) symmetry breaking
into the longitudinal component (Goldstone mode)
of the vector boson gauge coupling,
and does not need a Higgs field for its definition~\cite{CKM-remark}.
If a dilaton ${\cal D}$ descends from
the dynamical breaking of scale invariance,
its coupling to fermions $f$ and massive vector bosons $V$
should share a common dilution factor, $v/f$.
Could Nature have further subtleties involving the top quark?

\section{Discussion and Conclusion}
%

Before discussing strong Yukawa coupling further,
let us comment on flavor.
Extending to 4G naturally affects flavor physics,
such as $B_q \to \mu^+\mu^-$.
The combined analysis of CMS and LHCb has established~\cite{CMS:2014xfa}
$B_s \to \mu^+\mu^-$, albeit at $\sim 1\sigma$ below SM expectation.
More intriguing is $B_d \to \mu^+\mu^-$,
which has $3\sigma$ significance~\cite{CMS:2014xfa}, but only because
the central value is $4 \times$ 
SM!
This was our point~\cite{Hou:2013btm} in refuting
the verdict~\cite{Stone:2012yr} on 4G.
Experiments would surely pursue $B_d \to \mu^+\mu^-$,
and the larger its rate, the earlier the discovery~\cite{LHCb-Run-2}.
But it would need considerably more data
than at Run 1.

%
Another probe is the CPV phase $\phi_s$ in $B_s$-$\bar B_s$ mixing.
The measured $\phi_s = 
-0.030 \pm 0.033$~\cite{Vagnoni} from LHC Run~1,
which is dominated by LHCb, is fully consistent with SM, 
and further progress would take a few years.
But this just means $\vert V_{t^\prime s}^*V_{t^\prime b}\vert$
and $\arg(V_{t^\prime s}^*V_{t^\prime b})$ are small.
Keeping this constraint, we have shown~\cite{Hou:2014nna} that
enhanced $B_d \to \mu^+\mu^-$ can be accounted for,
while $K_L \to \pi^0\nu\bar\nu$ can be enhanced
up to the Grossman-Nir bound of $1.4 \times 10^{-9}$,
in correlation with some suppression of $B_s \to \mu^+\mu^-$.
These flavor and CPV probes would no doubt be pursued with vigor,
and could challenge the SM ``Higgs'' nature of the 125 GeV boson.
%
%
Lastly, one should not forget the baryon asymmetry of the Universe (BAU),
where the effective strength of CPV with 4G jumps by $10^{15}$~\cite{Hou:2008xd}
or more over 3G, and should suffice for BAU.
%
With such strong Yukawa coupling, one may have to rethink
the issue of the order of electroweak phase transition.

We return to discuss strong Yukawa before closing.

Just before Yukawa received the Nobel Prize, Fermi and Yang asked~\cite{FermiYang}
``Are mesons elementary particles?''
Defining ``elementary'' as ``structureless'',
they suggest the pion is an $N\bar N$ boundstate.
They could not treat, however,
the ultrarelativistic boundstate problem~\cite{UR-boundstate},
and the $\pi$-$N$ system took the path of QCD:
hadrons are stringy $q\bar q$ states.
But the well-known Goldberger-Treiman relation,
$\lambda_{\pi NN} \simeq \sqrt{2} m_N/f_\pi$, is of same form as Eq.~(5),
while the $g_{\pi NN}$ coupling extracted from $NN$ scattering
is of order $14$, the same strength as $\lambda_{\pi NN}$.
It was this NDA-strong coupling that made sense of Eq.~(6)
for 
the $G$-$Q$ system.
The situation is actually more crisp than the $\pi$-$N$ case:
$G$ is an exact NG boson, while $Q$, being sequential, is pointlike.
%
What would be the origin, or underlying theory, of such strong Yukawa couplings?
It must be as spectacular as QCD, but not a sequel,
 (``something new, and geometric''~\cite{CNYang}?),
hence not technicolor.
It is probably conformal~\cite{Goldberger:2008zz, Chacko, Csaki}.

Although the $G$-$Q$ system should not be stringy,
the similarity with the $\pi$-$N$ system, in particular
the NDA-strong coupling, suggests a simple analogy~\cite{Hou:2012df}
that may be of phenomenological relevance:
annihilation of $Q\bar Q \to nV_L$ into an
EW fireball of NG bosons $G$ (or $V_L$).
Fermi had already speculated about it,
but we learned since antiproton discovery
that $p\bar p$ annihilates at rest into a fireball of 5 pions on average,
emitted from a region of size $1/m_\pi$ at temperature $T \simeq 120$ MeV,
with Goldstone behavior of soft-pion suppression.
For the $Q\bar Q \to nV_L$ fireball~\cite{Hou:2012df},
one replaces $\pi \Rightarrow V_L$,
$1/m_\pi \Rightarrow 1/M_W$, and
$T$ slightly below EW phase transition temperature,
which together with $2m_Q$ determines the mean multiplicity
$n$ of ${\cal O}(10)$ or higher.
The Gaussian multiplicity distribution leads to
little impact on few boson final states.

Eq.~(6) implies $2m_Q = \sqrt2 \lambda_Q v \gtrsim 4$--$5$ TeV,
which seems out of reach at 14 TeV LHC,
and observing the fireball may not be easy.
But $Q\bar Q$ boundstates could help.
If the $V_L$ or NG boson $G$ is a massless $Q\bar Q$ boundstate (Fermi-Yang redux!),
the leading excitations~\cite{Enkhbat:2011vp} should be
$\pi_8$, $\omega_1$ and $\omega_8$~\cite{note},
where $G$ would be $\pi_1$ in this notation.
Although the hint for a 750 GeV $\gamma\gamma$ bump in
2015 data at 13 TeV has disappeared with more data~\cite{ICHEP2016},
it motivates one to consider how low in mass could these first excitations be.
For example, due to QCD repulsion, rather than attraction for $\pi_1/G$,
the $\pi_8$ color excitation could have mass below 1 TeV,
depending on its physical size.
But they would have to be produced in pairs.
Similar arguments can go for the less tightly bound $\omega_1$,
as compared with $\pi_1$, and its color excitation $\omega_8$,
which is of particularly interest, as it mixes with the gluon.
Unfortunately, their nonperturbative boundstate nature makes
the discussion rather speculative, as we have not solved the boundstate problem.
We note that the $\eta_1$, $\eta_8$ (as well as $\rho$ and $\sigma$) states
seem at best loosely bound~\cite{Enkhbat:2011vp} at $2m_Q$,
hence it would not have been easy to account for a 750 GeV $\gamma\gamma$ bump anyway.
But if low-lying boundstates exist around TeV, rather than 4--5 TeV,
the vector boson multiplicity of the fireball may be reduced,
and production may be aided by mixing of $\omega_8$ with the gluon.

In conclusion,
LHC Run 2 at 13 TeV 
would come to a close in 2018,
but New Physics is still no where in sight.
In face of Anderson's challenge that the 125 GeV boson itself
may not be in the SM Langrangian,
we have emphasized the possibility that it could still be the dilaton
arising from scale invariance violation of some conformal sector at high scale.
The SM ``Higgs'' nature of the 125 GeV boson should therefore  be scrutinized
free from any prejudice, and we must perform \emph{data-based},
simultaneous measurement of \emph{jet-tagged ggF and}
VBF production with LHC Run 2 data.
Heeding the cry, 
 ``\emph{Maybe the Higgs boson is fictitious!}'',
could turn out to be a second cross-fertilization from condensed matter physics.
If VBF is found suppressed, then the 125 GeV boson could be
a dilaton ${\cal D}$ rather than $H$,
with a heavy sequential quark doublet $Q$ as source of EWSB.
$Q\bar Q$ condensation by extremum-strength Yukawa coupling
implies $2m_Q \sim$ 4--5 TeV,
%
which could explain the absence of New Physics so far at the LHC,
motivating a higher energy collider.
But high multiplicity vector boson production might appear
at lower mass due to low-lying $Q\bar Q$ boundstates.
%
Corroborating evidence for a heavy sequential $Q$ could come from enhanced
rare decays such as $B_d \to \mu^+\mu^-$ and $K_L \to \pi^0\nu\bar\nu$.
Whether ascertaining VBF production or the pursuit of rare flavor physics,
the issue may take some years to pan out, but it could completely change
our perceptions of electroweak symmetry breaking.
While the above agenda needs to be checked,
the recent observation of $t\bar tH$ at expected strength in SM
poses a challenge that needs to be resolved.

\ 

\noindent{\bf Acknowledgement}.  Work supported in part by
MOST 103-2745-M-002-001-ASP and  NTU-EPR-105R8915.
We thank K.-F.~Chen, J.-M.~G\'erard, B. Grinstein,
T.~Han, J. Kubo, S.J.~Lee, M.~Lindner, S. Nussinov,
M.~Piai, A.~Soni, and especially S.~Paganis
and condensed matter colleagues
 C.-W.~Chang and C.-H.~Chern for useful discussions.
We are grateful to Y.~Mimura for consultations and encouragement.
%

\ 





\begin{thebibliography}{99}
\bibitem{farforward}
 Lack of resolution, if not detection ability,
 for very large $\eta$ makes it difficult 
 for measuring ``genuine'' VBF.
\bibitem{mired}
  This point is mired in the long language of Ref.~\cite{H-combo},
  as well as the individual ATLAS and CMS papers, and may
  not be easy to grasp. In any case, we maintain that one needs data-based
  simultaneous measurement of both VBF and jet-tagged ggF components.
\bibitem{int_range}
  From the point of view of solution finding,
  the large $\lambda_Q$ is not just strong coupling,
  but also for a more extended range of integration.
\bibitem{UR-boundstate}
  The massless $G$ as a Yukawa boundstate is astounding,
  and different from the pion.
  The $\gamma_5$ coupling nature requires large ``lower components'' of
  the Dirac spinors of $Q$ and $\bar Q$, hence ultra-relativistic motion.
  The total inertial energy, at several $2m_Q$
  (hence a rather compact object), would then have to be
  cancelled to zero by the ultrastrong Yukawa attraction.
\bibitem{dilaton}
  We make no claim for understanding what is the light ``Higgs mode''
  discovered far below the superconductor double-gap 2$\Delta$ in Ref.~\cite{squalid-H}.
  It should not be confused with our dilaton suggestion for 125 GeV boson.
\bibitem{CKM-remark}
  Hence, e.g. studies of CKM sector remains intact,
  as the ``Higgs boson'' usually does not appear
  in flavor processes, whether they are loop-induced or not.
\bibitem{note}
  We use hadron notation, as $Q$ is a doublet.
  With EWSB by $Q\bar Q$ condensation, there are colored partners.
  Note the absence of $\rho$-like states, in contrast to technicolor.
\end{thebibliography}
\end{document}